\documentclass[aip,amsmath,amssymb,reprint]{revtex4-1}

\usepackage{graphicx}
\usepackage{dcolumn}
\usepackage{bm}

\usepackage[utf8]{inputenc}
\usepackage[T1]{fontenc}
\usepackage{mathptmx}
\usepackage{etoolbox}
\usepackage{xcolor}

\usepackage{dotlessj}

\makeatletter
\def\@email#1#2{
 \endgroup
 \patchcmd{\titleblock@produce}
  {\frontmatter@RRAPformat}
  {\frontmatter@RRAPformat{\produce@RRAP{*#1\href{mailto:#2}{#2}}}\frontmatter@RRAPformat}
  {}{}
}
\makeatother

\begin{document}

\title[]{\textcolor{black}{How well can VMEC predict the initial saturation of external kink modes in near circular tokamaks and $l=2$ stellarators?}}

\author{R Ramasamy}
\affiliation{Max-Planck Institut f\"ur Plasmaphysik, Boltzmannstraße 2, 85748 Garching bei München, Germany}
\email{rohan.ramasamy@ipp.mpg.de}

\author{M Hoelzl}
\affiliation{Max-Planck Institut f\"ur Plasmaphysik, Boltzmannstraße 2, 85748 Garching bei München, Germany}

\author{S Henneberg}
\affiliation{Max-Planck Institut f\"ur Plasmaphysik, Wendelsteinstrasse 1, 17491 Greifswald, Germany}

\author{E Strumberger}
\affiliation{Max-Planck Institut f\"ur Plasmaphysik, Boltzmannstraße 2, 85748 Garching bei München, Germany}

\author{K Lackner}
\affiliation{Max-Planck Institut f\"ur Plasmaphysik, Boltzmannstraße 2, 85748 Garching bei München, Germany}

\author{S G\"unter} 
\affiliation{Max-Planck Institut f\"ur Plasmaphysik, Boltzmannstraße 2, 85748 Garching bei München, Germany}


\begin{abstract}
      The equilibrium code, VMEC, is used to study external kinks in low $\beta$ tokamaks and $l=2$ stellarators. The applicability of the code when modelling nonlinear MHD effects is explored in an attempt to understand and predict how the \textcolor{black}{initial saturation of the MHD mode} depends on the external rotational transform. It is shown that \textcolor{black}{helicity preserving, free boundary} VMEC computations do not converge to a single perturbed solution with increasing spectral resolution. \textcolor{black}{Additional constraints are therefore applied to narrow down the numerical resolution parameters appropriate for physical scans}. The dependence of the modelled (4, 1) kink mode on the external rotational transform and field periodicity is then studied. \textcolor{black}{While} saturated states can be identified which decrease in amplitude with increasing external rotational transform\textcolor{black}{, bifurcated states are found that contradict this trend. It was therefore not possible to use VMEC alone to identify the physical dependency of the nonlinear mode amplitude on the magnetic geometry. The accuracy of the VMEC solutions is nevertheless demonstrated by showing that the expected toroidal mode coupling is captured in the magnetic energy spectrum for stellarator cases.} Comparing with the initial value code, JOREK, the predicted redistribution of poloidal magnetic energy from the vacuum to plasma region in VMEC is shown to be physical. This work is a first step towards using VMEC to study MHD modes \textcolor{black}{in stellarator geometry}.
\end{abstract}


\maketitle

\section{Introduction} \label{sec:intro}
The VMEC free-boundary code \cite{HirshmanRij1986} has been proposed as a computationally efficient solution to nonlinear MHD problems in the ideal MHD limit. Though VMEC and other equilibrium codes do not evolve the MHD equations dynamically, it has been proposed that perturbed equilibrium states can be found which correspond to the final saturated state of a MHD perturbation\cite{cooper2010tokamak}. This is considered possible if the saturated state, referred to as the perturbed solution, is appropriately constrained to have a physically meaningful link to the initial unperturbed equilibrium from which the dynamics evolve \cite{turnbull2012plasma, turnbull2013comparisons}.

\textcolor{black}{Equilibrium codes have been used successfully to} study externally driven problems, such as resonant magnetic perturbations (RMPs) \cite{chapman2014three, king2015experimental}, as well as to study ideal MHD instabilities in a variety of tokamak scenarios \cite{graves2012magnetohydrodynamic, strumberger2014mhd, cooper2016saturated, kleiner2019current}. Perturbed solutions of both internal, and external ideal MHD modes have been validated against other linear, and nonlinear MHD approaches\cite{brunetti2014ideal, ramasamy2022}. \textcolor{black}{In many of these studies, VMEC is used to analyse MHD instabilities that have been observed experimentally, such that there is a strong understanding of what the expected perturbed solution should be.}

Regarding stellarators, only a few studies of \textcolor{black}{the nonlinear saturation of MHD modes} using such an equilibrium approach exist\cite{cooper2018stellarator, garabedian2006three}. In Ref. \onlinecite{cooper2018stellarator}, periodicity breaking modes in a HELIAS configuration were found to lead to much milder edge distortions than those observed in similar tokamak studies, even at high plasma $\beta$ -- the ratio of plasma and magnetic pressure. \textcolor{black}{This is one of the few applications in the literature, where it has been attempted to use VMEC to predict the ideal saturation of a MHD mode that has not been experimentally observed}. Though these results did not conserve helicity, or enforce physical constraints that link the unperturbed equilibria and saturated states explicitly, \textcolor{black}{they have motivated discussion regarding the use of equilibrium codes to understand nonlinear MHD phenomena in stellarators, informing their design and optimisation.} 

\textcolor{black}{While the above results are encouraging, it is known that VMEC can produce apparently unphysical results when compared with linear MHD codes\cite{turnbull2013comparisons}. Rigorous numerical resolution scans from RMP studies also identify that current-preserving, free boundary VMEC computations do not converge to a single well-defined solution for externally driven problems\cite{lopez2020validation}.}

\textcolor{black}{If VMEC is to be applied to optimise and design stellarators for good nonlinear MHD properties, a rigorous understanding of the applicability of the equilibrium approach and its limits in modelling nonlinear MHD activity must be obtained. It is known that fixing the $\iota$ profile provides a physical link between the initial and perturbed solutions as the conservation of helicity is enforced. While this constraint is necessary, it is not sufficient to demonstrate that the obtained numerical equilibria are physically meaningful. Additional arguments based on linear and nonlinear simulation results, as well as experimental comparisons are often needed in order to defend the results of the equilibrium approach.}

\textcolor{black}{The main ambition of this work is to understand how far VMEC can be used to study \textcolor{black}{the nonlinear saturation of MHD modes}, relying as little as possible on other more expensive nonlinear computations or experimental observations to justify results. In such a way, the authors would like to understand to what extent VMEC can be used to predict the nonlinear saturation, rather than act as a diagnostic that complements more expensive numerical or experimental studies.}

In this paper, nonlinearly perturbed states are computed for tokamaks, and $l=2$ stellarators with near circular, elliptical cross section. While these devices are much simpler than the optimised configurations that are of interest to the fusion research community at the moment, they provide a reasonable test bed for understanding the numerical methods involved, and how to appropriately apply them. \textcolor{black}{This study also aims to explore a physics question --- how does the saturated amplitude, and the nonlinear mode structure of ideal MHD instabilities vary as a function of the applied external rotational transform and field periodicity, denoted by $\iota_\text{ext}$ and $N_\text{p}$, respectively?} 

The remainder of this paper is outlined as follows. The method for finding nonlinearly perturbed states in VMEC, and the parameter space in which free boundary perturbed VMEC equilibria are computed is discussed in Section \ref{sec:vmec_equilibra}. In Section \ref{sec:res_scans}, the resolution requirements are studied, identifying that the saturation amplitude of the perturbed state can depend on the spectral resolution of the computation, and that current sheets form at the plasma boundary. These results are consistent with previous studies that suggest the saturation of an ideal kink instability requires some localised resistivity to prevent current sheets from growing indefinitely large \cite{arber1999unstable}. 

\textcolor{black}{The numerical convergence properties presented in Section \ref{sec:res_scans} introduce a challenge for resolving the saturated mode. Multiple, equally valid solutions for the saturated MHD perturbation exist --- how does one differentiate between these solutions to find the most physical one?} Further constraints, such as incompressibility, which is an expected condition for high aspect ratio external kink modes \cite{kadomtsev1973}, and the increase in toroidal plasma current --- the $I_p$ spike --- from the initial to perturbed equilibrium are used to identify the resolution parameters which lead to the most physically meaningful saturated mode. \textcolor{black}{The Fourier content of induced current sheets at the plasma boundary is also considered as a constraint on the perturbed solutions.} 

The above indicators are used \textit{a posteriori}, after the VMEC computations are converged, to narrow down the numerical parameter space in VMEC sufficiently to proceed with scans of the physical parameters governing the external kink mode --- namely, the $q$ profile and $\iota_\text{ext}$. In Section \ref{sec:iota_q_scan}, the dependence of the nonlinearly saturated mode structure on these parameters is evaluated. In most results, the nonlinear perturbation amplitude follows the linear growth rate of the initial ideal MHD instability. However, \textcolor{black}{within the narrow window of numerical parameters that are considered physically reasonable,} deviations are also observed in some solutions which are thought to be alternative bifurcated states of the equilibrium. \textcolor{black}{In such a way, while physically meaningful saturated states could be obtained, the authors were unable to identify the physical dependencies of the mode conclusively using VMEC alone.}

\textcolor{black}{To further interrogate the physical accuracy of the modes observed in VMEC, the nonlinear magnetic energy spectrum is considered in Section \ref{sec:nonlinear_energies}. The expected mode coupling of the MHD instability is identified in the stellarator cases}. It is shown that the stellarator perturbations are dominated by toroidal harmonics that are within the toroidal mode family of the instability. In addition, it is shown that the $n=0$ magnetic energy inside the simulation domain increases, despite being the dominant contributor to the instability.

\textcolor{black}{The analysis in Section \ref{sec:nonlinear_energies} presents us with a further question --- why does the $n=0$ magnetic energy increase?} To answer this question, observations from VMEC computations are compared with those from a free boundary tokamak simulation using the initial value code, JOREK, in Section \ref{sec:jorek_comp}. The results of the initial value code are used to demonstrate that the increase in the $n=0$ magnetic energy in VMEC is physically reasonable. \textcolor{black}{In such a way, VMEC captures many aspects of the energy dynamics accurately.} The implications of the results of this study are discussed in Section \ref{sec:conclusion} with an outlook for future work.

\section{Computation of perturbed free boundary VMEC equilibria} \label{sec:vmec_equilibra}

The algorithm implemented in VMEC is well documented in Ref. \onlinecite{hirshman1983steepest}, and the methods applied in this paper follow the approach of previous studies, which have used VMEC to compute nonlinearly perturbed equilibrium states \cite{strumberger2014mhd, kleiner2019current}. A brief review of this approach follows in this Section. 

The ideal MHD potential energy, $W_\text{mhd}$, can be written as

\begin{equation} \label{eq:mhd_energy}
    W_\text{mhd} = \int \frac{B^2}{2 \mu_0} + \frac{p}{\gamma - 1}\ dV.
\end{equation}

where $B$ is the magnetic field strength, $p$ is the plasma pressure, and $\mu_0$ and $\gamma$ are the vacuum magnetic permeability, and heat capacity ratio, respectively. VMEC minimises $W_\text{mhd}$ of the plasma and vacuum region up to a prescribed level of accuracy in the ideal MHD force balance equation

\begin{equation} \label{eq:force_residual}
    \mathbf{j} \times \mathbf{B} - \nabla p = 0,
\end{equation}

where $\mathbf{j}$, $\mathbf{B}$ and $p$ are the current, magnetic field and plasma pressure, respectively.

The MHD energy can be further minimised by physical MHD perturbations if the initially targeted equilibrium is ideal MHD unstable. In this case, a new equilibrium is obtained, which is physically interpreted as \textcolor{black}{the nonlinearly saturated MHD state produced by the ideal MHD equations, which govern the dynamics. It is important to acknowledge that the hydromagnetostatic equations used in VMEC cannot represent the instantaneous dynamics of the MHD mode, during its growth and saturation. The outcome of the overall dynamics can however be studied by considering the change in equilibrium quantities between the initial and saturated states, as attempted in Section \ref{sec:nonlinear_energies}.} 

The interpretation of the perturbed equilibrium as a saturated state can only be justified for the study of the ideal saturation of MHD modes, such that effects leading to internal reconnection, and the breaking of the internal nested flux surfaces in VMEC can be neglected. For this reason, the focus of this study is on the initial ideal saturation of the external kink. 

When using VMEC to study the long term behaviour of MHD instabilities, the applicability of the method comes into question. A convincing argument for the long term stability of the perturbed solution is necessary to justify these applications. Such arguments can be based on the use of external plasma control, such as in the case of RMPs, or analysis to show the perturbed solution is stable to further non-ideal MHD perturbations.


To study external modes using VMEC, the plasma boundary must be allowed to deform. This requires a free boundary solution, which in turn requires a representation of the externally generated vacuum magnetic field. The virtual casing principle \cite{shafranov1972use}, which has been implemented in EXTENDER \cite{drevlak2005pies}, can be used together with the NESCOIL \cite{merkel1987solution} and MAKEGRID codes to compute the magnetic field from external currents. \textcolor{black}{All VMEC results presented in this study are free boundary, using this technique to generate the vacuum field.}

\begin{figure*}
    \centering
    \centering
    \includegraphics[width=0.99\textwidth]{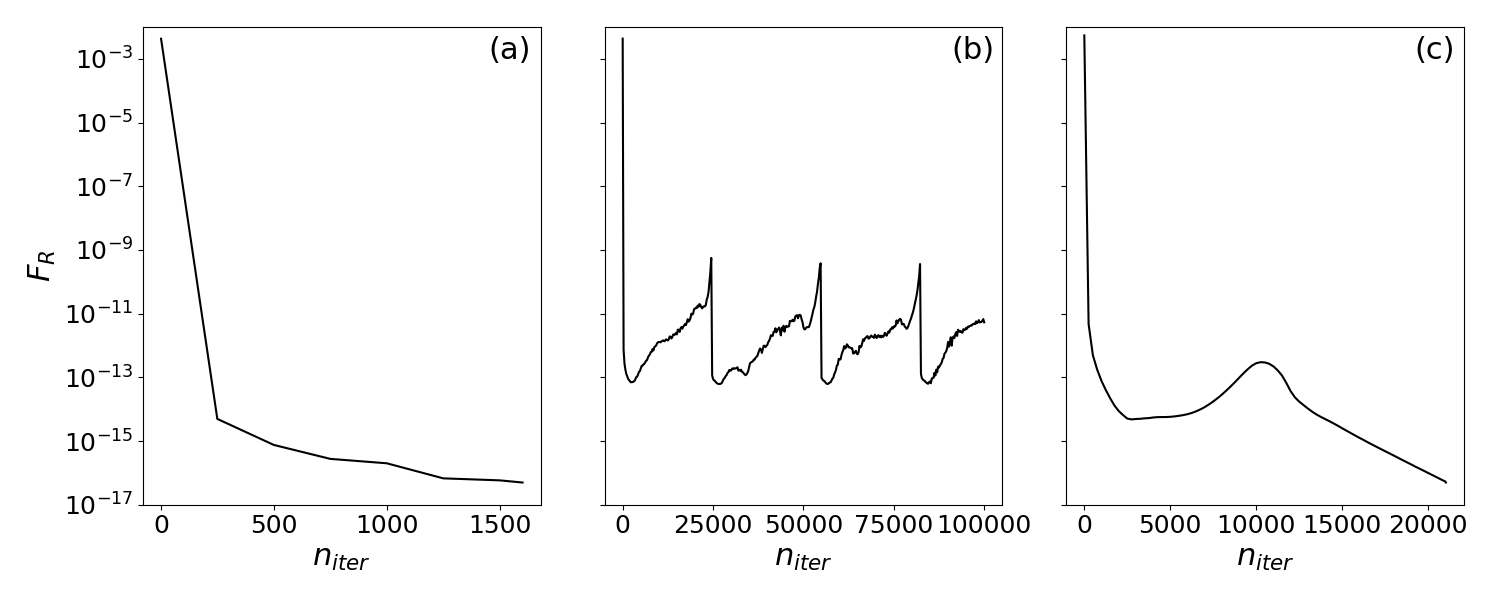}
    \caption{Plots of the force balance residual, $F_R$, for VMEC computations with different convergence behaviours. An unperturbed case (a) and perturbed cases with poor (b) and good (c) convergence are shown. \textcolor{black}{All perturbed states reported in this paper have the convergence behaviour shown in Figure \ref{fig:vmec_convergence} (c).}}
    \label{fig:vmec_convergence}
\end{figure*}


As mentioned in Section \ref{sec:intro}, helicity conservation is an appropriate constraint for studying ideal MHD perturbations. For this reason, \textcolor{black}{the perturbed equilibria studied in this paper have been computed fixing the $\iota$ profile to the form of the initial equilibrium during the computation}. This enforces that the helicity of the plasma is conserved\cite{turnbull2012plasma, ramasamy2022}. Finding perturbed states that constrain $\iota$ instead of the toroidal current density $j$ can be more challenging, but is necessary to ensure a physical link between the initial and perturbed equilibrium state.

\subsection{VMEC convergence behaviour} \label{sec:vmec_convergence}

The different convergence behaviours of the radial force residual, $F_R$, in VMEC are shown in Figure \ref{fig:vmec_convergence}. In this work, the tolerance in the force residuals for convergence, $f_\text{tol}$, is set to be less than $10^{-16}$. Figure \ref{fig:vmec_convergence} (a) shows an equilibrium that converges successfully to an unperturbed solution. In this unperturbed case, the force residual decreases monotonically, as is expected with the minimisation of the $W_\text{mhd}$ energy. 

For the perturbed cases in Figure \ref{fig:vmec_convergence} (b) and (c), it can be seen that the force residuals increase, as the instability is found and the force balance of the equilibrium is lost. Note that this often occurs at a relatively low residual value, below $10^{-13}$, which would normally be considered reasonably converged in equilibrium studies. If a new perturbed state is found, the residuals must fall as the new perturbed state is reached, leading to the behaviour in Figure \ref{fig:vmec_convergence} \textcolor{black}{(c)}. Similar observations of the force residual have been observed in studies of magnetic island formation using NSTAB \cite{garabedian2006three}.

In many cases however, the force residuals continue to rise, until the VMEC algorithm detects the unfavourable behaviour. The equilibrium is then reset to an earlier stage, and the numerical time step parameter is decreased in an attempt to improve convergence. Often the smaller numerical time step does not lead to a converged solution and the process repeats indefinitely, as shown in Figure \ref{fig:vmec_convergence} (b). Though the residuals still remain relatively low compared to normal equilibrium studies, the results produced with this behaviour are either considered to be numerical, or disruptive MHD instabilities that cannot be modelled with VMEC. The perturbed states reported in this paper all have the convergence behaviour shown in Figure \ref{fig:vmec_convergence} (c).

\subsection{Equilibria and scale invariance}

In this study, simple tokamaks and classical $l=2$ stellarators were targeted at $\beta \approx 0$. As a result, due to the scale invariance of ideal MHD, the behaviour of the saturated modes can be assumed to be a function of the magnetic geometry alone. This means that the parameter space at $\beta\approx 0$ is defined by the external rotational transform, $\iota_\text{ext}$, and the safety factor profile, $q(s)$, where $q(s)$ is in turn dependent on $j_0(s)$ and the shaping of the plasma.

To simplify the parameter space further, the initial toroidal current density, $j_0(s)$, of the considered equilibria is prescribed to have an Ohmic profile, similar to the analytic studies by Wesson \cite{wesson1978hydromagnetic}

\begin{equation} \label{eq:j_0}
    j_0(s) = 1 - s^p,
\end{equation}

where $s$ is the normalised toroidal flux, $\hat \Phi$, and $p$ is a constant. \textcolor{black}{For the final results shown in this paper, which constrain $\iota(s)$, the rotational transform is first determined by computing an unperturbed equilibrium with the current profile prescribed by equation \ref{eq:j_0}.}

\begin{figure}
    \centering
    \includegraphics[width=0.4\textwidth]{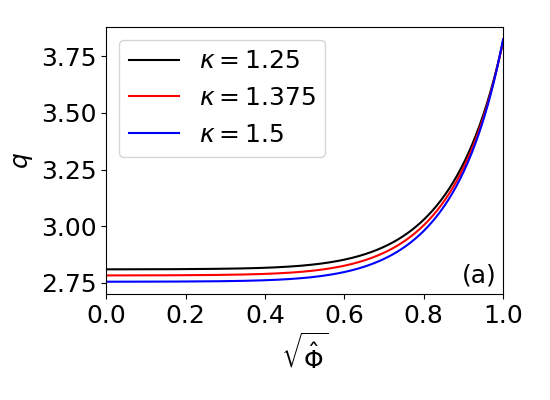}
    
    \includegraphics[width=0.49\textwidth]{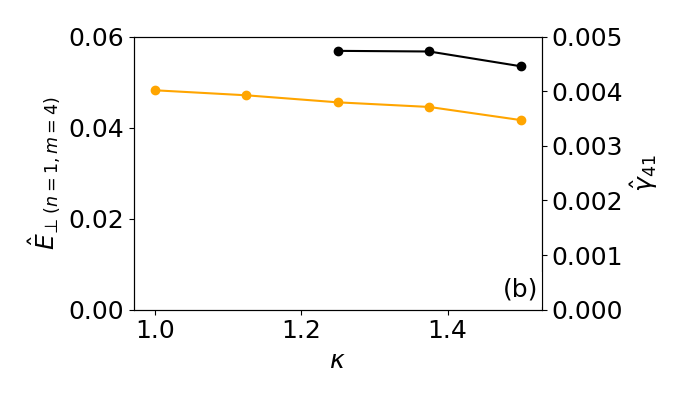}
    \caption{Plots of q profile (a) and normalised displacement amplitude (b) of the perturbed solution for tokamak cases with $p=3$, $q_\text{edge}=3.85$, varying $\kappa$. The displacement amplitude (b, black, left axis) is compared with the normalised growth rate (b, orange, right axis) computed with CASTOR3D. Perturbed states with good convergence properties were found only for $\kappa >= 1.25$.}
    \label{fig:kappa_scan}
\end{figure}

All equilibria considered have a high aspect ratio, $A=a/R_0=0.1$, where $a$ is the minor radius, and $R_0$ is the major radius. In order to determine a parameter space for finding nonlinearly perturbed states, tokamak cases were first considered, using $p=2-4$. The elongation, $\kappa=b/a$, where $b$ is the maximum height of the toroidally averaged plasma boundary, defined by the $n=0$ boundary coefficients, above the major axis, and edge safety factor, $q_\text{edge}$ were also varied. 

Equilibria initialised with quartic current profiles were difficult to converge to a saturated state, while constraining the $q$ profile. Using cubic and quadratic profiles, saturated (4, 1) modes could be found. Figure \ref{fig:kappa_scan} (a) shows the variation of the $q$ profile with the plasma elongation for simulations with cubic current profiles, setting $q_\text{edge} = 3.85$. The saturated state could only be found in VMEC for $\kappa \ge 1.25$. For lower values approaching a circular plasma, the convergence behaviour in Figure \ref{fig:vmec_convergence} \textcolor{black}{(b)} was observed. 

\begin{figure*}
    \centering
    \includegraphics[width=0.99\textwidth]{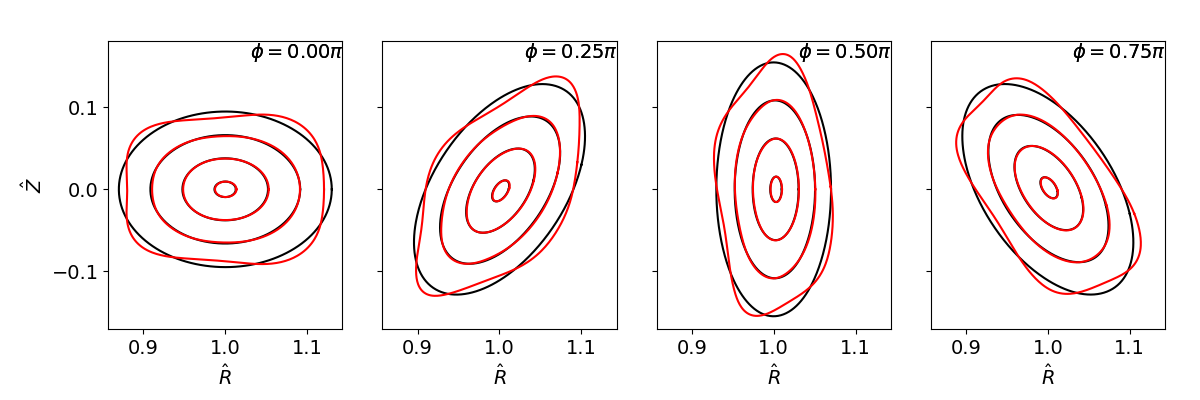}
    \caption{\textcolor{black}{Equilibrium flux surfaces of the initial (black) and perturbed (red) solutions of a $l=2$, $N_p=2$ stellarator case with $\iota_\text{ext}=0.098$. The shape of the initial equilibrium is modified from a typical rotating ellipse by vertical shaping, such that $\kappa = 1.25$. A (4, 1) external kink perturbation is visible in the perturbed solution. $\hat R$ and $\hat Z$ are normalised by $R_0$.}}
    \label{fig:l2_stellarator_surfaces}
\end{figure*}

The normalised growth rate of the (4, 1) external kink mode, $\hat \gamma_{41}=\gamma/\tau_A$, can be computed from the unperturbed equilibrium state using CASTOR3D \cite{Strumberger2016}. The growth rate is normalised by the Alfv\'en time, $\tau_A = B / \sqrt{\mu_0 \rho}$, where $\rho$ is the ion mass density. This accounts for variations in the magnetic field strength between equilibria. The growth rates are shown in Figure \ref{fig:kappa_scan} (b). It can be seen that $\hat \gamma_{41}$ increases with decreasing $\kappa$. As such, it is thought that the poor convergence at low $\kappa$ is due to the instability being too strong to reach a saturated state. Quadratic profiles were also computed, but again the saturated states did not converge well when decreasing $\kappa$. Scans with a higher $q_\text{edge}$, targeting a (5, 1) instability also failed to converge for plasmas with lower elongation. As a result of the above analysis, the tokamak and stellarator equilibria studied herein all have $\kappa=1.25$.  The magnetic geometry is modified by varying $q(s)$ and $\iota_\text{ext}$. 

\textcolor{black}{An example of the initial and perturbed equilibria for the $l=2$, $N_p=2$ stellarator case with $\iota_\text{ext}=0.098$ is shown in Figure \ref{fig:l2_stellarator_surfaces}. The imposed vertical shaping to get $\kappa=1.25$ has modified the initial equilibrium from the typical rotating ellipse expected for $l=2$ stellarators. When compared with the initial equilibrium, the perturbed solution has a clear (4, 1) external kink deformation. It has been confirmed that all perturbed solutions presented in this work are dominated by such (4, 1) kink perturbations, as expected.}

\section{Resolution scans} \label{sec:res_scans}
Before proceeding to scans of the physical parameters, $q_\text{edge}$ and $\iota_\text{ext}$, it is important to understand the influence of resolution parameters on the perturbed equilibria that are being considered. Resolution scans of the normalised plasma boundary displacement, plasma current spike and plasma incompressibility are carried out in Section \ref{sec:glob_quantities}. Analysis of these global quantities helps to isolate the numerical parameter space where physically meaningful solutions are expected. In Section \ref{sec:current_sheets}, the dependence of localised current sheets at the plasma boundary on numerical parameters is studied.

\subsection{Analysis of global quantities} \label{sec:glob_quantities}

\begin{figure}
    \centering
    \includegraphics[width=0.49\textwidth]{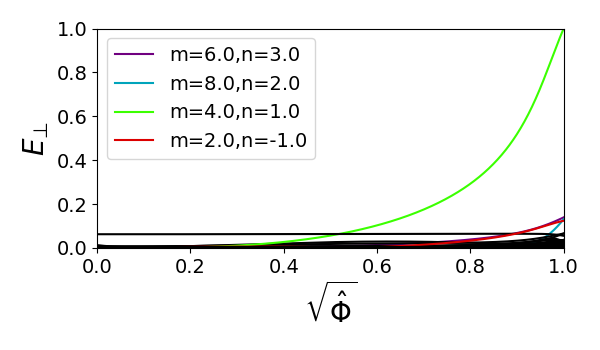}
    \caption{Fourier decomposed nonlinearly perturbed displacement of the VMEC solution for a $l=2$ stellarator with two fold periodicity and $\iota_\text{ext}=0.098$.}
    \label{fig:fourier_eigenfunction_comp}
\end{figure}

The normalised Fourier decomposed, perpendicular plasma displacement, $\hat E_\bot$, is computed in straight field line coordinates using a similar method to previous studies \cite{kleiner2018free, strumberger2014mhd}. Figure \ref{fig:fourier_eigenfunction_comp} shows $\hat E_\bot$ for a stellarator case with $q_\text{edge}=3.9$, $\iota_\text{ext}=0.098$, and two fold periodicity. It can be seen that the Fourier displacement is dominated by the $n=1$, $m=4$ Fourier component. The poloidal and toroidal sidebands are below $\approx 10\ \%$ of the dominant mode, as expected for an instability in a low $\beta$, high aspect ratio $l=2$ stellarator, where poloidal and toroidal mode coupling should be limited. As a result, the $n=1$, $m=4$ Fourier component alone, $E_{\bot, (n=1,\ m=4)}$, is used as a measure of the mode amplitude and its variation with resolution parameters.

\begin{figure*}
    \centering
    \includegraphics[width=0.9\textwidth]{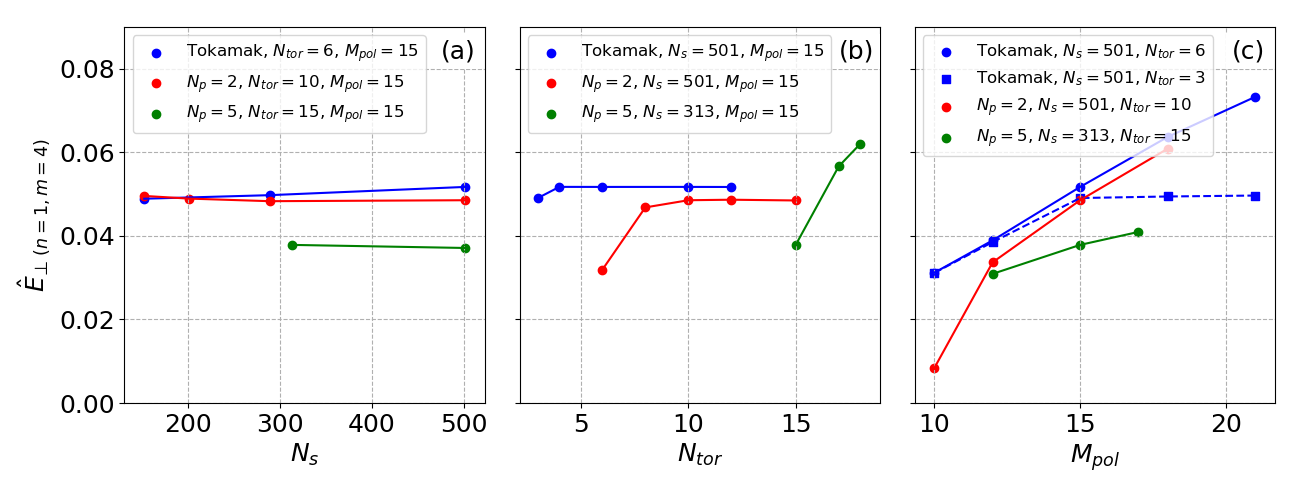}
    
    \includegraphics[width=0.9\textwidth]{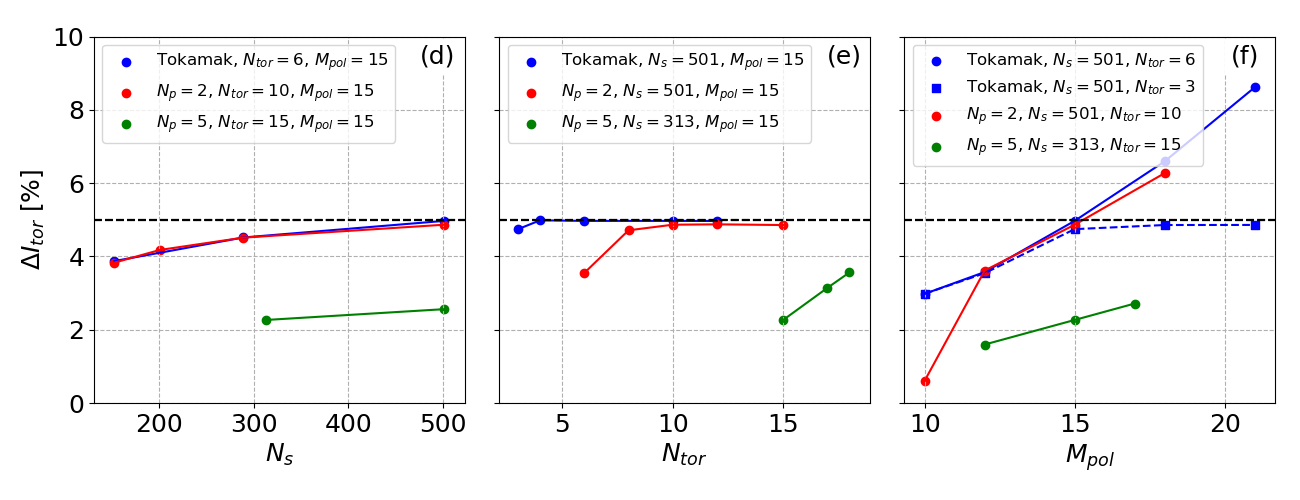}
    
    \includegraphics[width=0.9\textwidth]{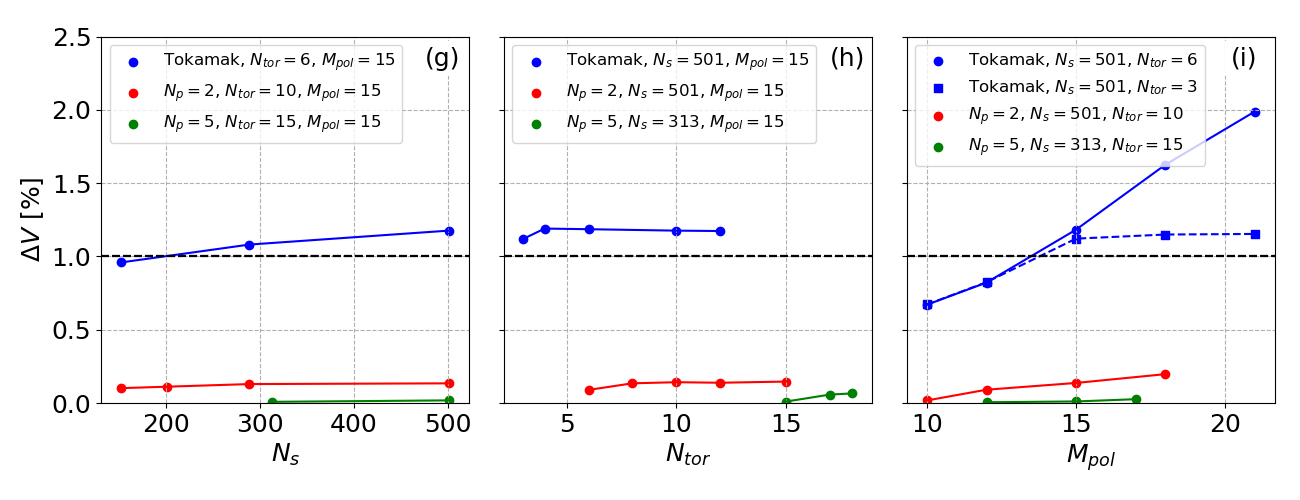}    
    
    \caption{Resolutions scans for the boundary displacement (a-c), plasma current spike (d-f) and change in plasma volume (g-i) as a function of $N_s$ (a,d,g), $M_\text{pol}$ (b,e,h), and $N_\text{tor}$ (c,f,i) for tokamak (blue), $N_p=2$ (red) and $N_p=5$ (green) stellarator cases with $q_\text{edge}=3.9$. The stellarator cases both have $\iota_\text{ext}\approx 0.1$. \textcolor{black}{The expected $5\ \%$ current spike and $1\ \%$ upper bound for the change in plasma volume are marked by black horizontal dashed lines in (d-f) and (g-i), respectively.}}
    \label{fig:vmec_kink_resolution_scan}
\end{figure*}

Figure \ref{fig:vmec_kink_resolution_scan} (a-c) show $\hat  E_{\bot, (n=1,\ m=4)}$ as a function of the radial, $N_s$, toroidal, $N_\text{tor}$ and poloidal, $M_\text{pol}$, resolution in VMEC. \textcolor{black}{It is important to note that, for stellarator cases, the modelled (4, 1) external kink mode in this study breaks the periodicity of the device. In order to observe this mode, it is necessary to include all toroidal harmonics in the computation up to the maximum harmonic, $N_\text{tor}$.} 

Results are shown for a tokamak and $l=2$ stellarators with $N_p=2$ and $5$ with $q_\text{edge}=3.9$ and $\iota_\text{ext}\approx0.1$ in the stellarator cases. Similar to previous studies of RMPs \cite{lopez2020validation}, it is found that computations do not converge to a single solution with increasing spectral resolution. While the displacement of the plasma boundary is converged at reasonable radial resolutions, the poloidal resolution is not converged for any of the simulated cases. It should be noted that higher resolution simulations were attempted but were not tractable due to high computational cost, leading to convergence difficulties. 

In Ref.\onlinecite{lopez2020validation}, it was argued that this could be due to the high poloidal resolution required in order to resolve the resonant poloidal mode number at the edge. The implied contributions to the saturated mode structure from the higher poloidal harmonics is plausible, given that $m=4n$ modes are required to resolve the resonant poloidal mode of a given toroidal mode number. In order to confirm that the solution converges when the resonant harmonic is included, computations using $n=3$ are shown in Figure \ref{fig:vmec_kink_resolution_scan} (c). At this lower toroidal resolution, the computation converges at approximately $M_\text{pol}=15$, soon after the resonant $m=12$ mode is passed. 


In addition to this observation, the toroidal resolution requirements increase with $N_p$, such that edge displacements could not be converged for the $N_p=5$ case. This is likely due to higher toroidal components of the mode as a result of the mode coupling produced by the $n>0$ components of the background vacuum magnetic field.

The above observations present a challenge to the use of the VMEC equilibrium approach for physics studies --- all of the above equilibria are valid perturbed equilibrium states that could correspond to a saturated ideal MHD mode, so how does one determine the numerical resolution which is most representative of the saturated state produced by the ideal MHD dynamics? 

To address this question, additional physical constraints need to be applied to the equilibrium solution. In this study, additional constraints have not been added to the convergence algorithm in VMEC; rather, they are used to determine the converged equilibria which are the most physically relevant. In such a way, these constraints are not used to modify the VMEC convergence algorithm. Instead, the adherence of converged computations at different resolutions to these constraints is assessed \textit{a posteriori}, as shown in Figure \ref{fig:vmec_kink_resolution_scan}. Two possible constraints were explored, using the change in the toroidal plasma current and plasma volume, as shown in Figure \ref{fig:vmec_kink_resolution_scan} (d-f) and (g-i), respectively.

The first constraint considered is to ensure that the observed plasma current spike, typically observed during ideal MHD activity, between the perturbed and unperturbed solutions is reasonable in magnitude. \textcolor{black}{Given the lack of experimental data to inform this constraint, a current spike on the order of $5\ \%$  at $q_\text{edge}=3.9$ is assumed to be reasonable.} The variation of the plasma current spike observed in resolution scans is shown in Figure \ref{fig:vmec_kink_resolution_scan} (d-f). It can be seen that the current spike remains below $5\ \%$ for the tokamak and $N_p=2$ case when $M_\text{pol} <= 15$, such that results below this value are taken to be more physically meaningful.

A second constraint can be applied by noting the conserved quantities that are found in conventional high aspect ratio orderings of the dynamics in fusion devices. In particular, the incompressibility of the plasma is typically preserved to second order in the aspect ratio \cite{kadomtsev1973}. As a result for the simulated cases with $A=0.1$, it is expected that the plasma volume will be conserved up to $\approx 1\%$. The change in plasma volume is shown in Figure \ref{fig:vmec_kink_resolution_scan} (g-i). For the stellarator cases, the change in plasma volume is relatively small. In the tokamak case, \textcolor{black}{the change in plasma volume can exceed $1\%$}, with a strong dependence on the poloidal resolution for $M_\text{pol}\ge15$. For this reason, it could be argued that solutions with a lower poloidal resolution are more reasonable.

\subsection{Structure of edge current sheets} \label{sec:current_sheets}

For an ideal external kink mode, current sheets are expected to form during the dynamics, wrapping around the outside of the main plasma column at the unstable resonant surface\cite{arber1999unstable}. Further, the theory in Ref. \onlinecite{kadomtsev1973} predicts that current sheets are also induced at the plasma boundary as a response to its deformation on the ideal MHD timescale. In such a way, large localised current sheets are expected in the perturbed equilibrium solution, according to the physical model implemented in VMEC.

\begin{figure}
    \centering
    \includegraphics[width=0.45\textwidth]{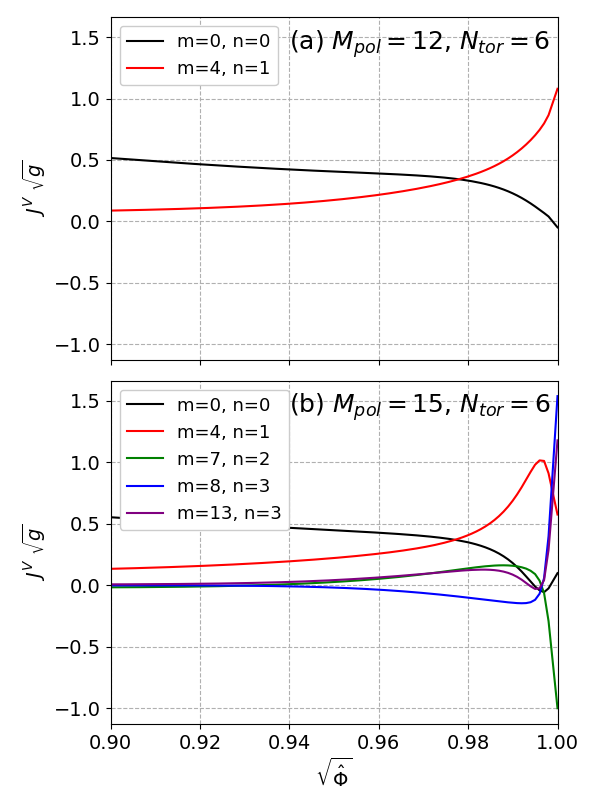}
    \caption{Fourier components of the toroidal current density profile near the plasma boundary for the tokamak case studied in Figure \ref{fig:vmec_kink_resolution_scan}, using different spectral resolutions. $J^V \sqrt{g}$ is normalised to the value of the $m=0$, $n=0$ Fourier component at the magnetic axis. For the $M_\text{pol}=12$, $N_\text{tor}=6$ case (a), the $m=4$, $n=1$ Fourier component is dominant at the plasma edge. The $M_\text{pol}=15$ case (b) contains much broader Fourier content, including components resonant on internal rational surfaces and some modes which could be numerical artifacts.}
    \label{fig:vmec_tok_fixed_radial_scan_current_sheets}
\end{figure}

The growth of these current sheets, and the kink mode more generally, would normally be limited by non-ideal effects such as resistive diffusion. The omission of non-ideal effects in VMEC could be the reason that the saturated kink amplitude is observed to increase indefinitely with increasing spectral resolution. \textcolor{black}{To help identify the most physically meaningful solution, it would be instructive to understand both the expected magnitude of the current sheets, and what their expected structure should be. Similar to Ref. \onlinecite{lazerson2016verification}, the Fourier decomposed toroidal current density can be plotted near the plasma boundary, as shown in Figure \ref{fig:vmec_tok_fixed_radial_scan_current_sheets}. The Fourier contributions are normalised by the value of the $m=0$, $n=0$ component on the magnetic axis, and only the dominant Fourier components have been plotted.}

\textcolor{black}{There are qualitative differences in the current sheets observed in the perturbed solutions using the different spectral resolution parameters in Figure \ref{fig:vmec_kink_resolution_scan}. For the $M_\text{pol}=12$ tokamak case, shown in Figure \ref{fig:vmec_tok_fixed_radial_scan_current_sheets} (a), the ($m=4$, $n=1$) component of the current is large at the plasma edge, as one might expect for the observed instability. In the $M_\text{pol}=15$ tokamak case, shown in Figure \ref{fig:vmec_tok_fixed_radial_scan_current_sheets} (b), the Fourier content of the current sheets is broader. Some of the additional components could correspond to a nonlinear response of internal resonant surfaces. For example, the $m=7$, $n=2$ mode could correspond to shielding currents at the $q=3.5$ surface, which prevent internal reconnection within the plasma core.}

\textcolor{black}{Even if this is the case, it is unclear whether these shielding currents should be considered physical. In Appendix \ref{sec:jorek_currents}, it is shown that internal (7, 2) magnetic islands can form near the plasma edge, when resistive effects are included in the nonlinear phase. It makes sense that in VMEC, the (7, 2) islands would be shielded if the resistive effects in the plasma edge region are sufficiently small, or neglected --- a solution similar to Figure \ref{fig:vmec_tok_fixed_radial_scan_current_sheets} (b) may not be precluded in this limit.}

\textcolor{black}{There are also unintuitive contributions to the induced current in Figure \ref{fig:vmec_tok_fixed_radial_scan_current_sheets} (b). In particular, the $m=8$, $n=3$ mode, which does not have a corresponding resonance within the plasma or vacuum region, is unexpected. This mode is larger than the $m=4$, $n=1$ mode at the plasma boundary. As a physical explanation for the dominance of this contribution could not be found, it could be a numerical artifact.} 

\textcolor{black}{A rigorous method for determining whether the structure of the current sheets in the VMEC solution are correct remains an open question. If the solutions in Figure \ref{fig:vmec_tok_fixed_radial_scan_current_sheets} are to be differentiated in terms of their physical validity, a more detailed physical intuition of the nonlinear mode structure of the shielding currents is required. Developing this intuition is beyond the scope of the current work.}

\subsection{Choice of resolution parameters} 

While the above analysis cannot provide a definitive answer for what the correct resolution parameters are to obtain the most physically meaningful perturbations, the change in volume and current spike imply that poloidal resolutions below $M_\text{pol}=15$ are likely to be most reasonable. In the scans of physical parameters which follow, $M_\text{pol}=15$, and $N_s \ge 300$ are used, unless stated otherwise. Although the current sheets for the $M_\text{pol}=15$ case may have unphysical numerical structures, this value was chosen because, for the scan of $q_\text{edge}$ in Section \ref{sec:q_scan}, it was difficult to obtain saturated states close to the upper and lower linear stability thresholds for $M_\text{pol} < 15$. This resolution was therefore necessary in order to explore the physical parameter space of the mode in Section \ref{sec:q_scan}. The toroidal resolution is $N_\text{tor}=6$ and $N_\text{tor}=15$ for the tokamak and $N_p=5$ cases, respectively. Results with different toroidal resolution are computed for the $N_p=2$ case, to investigate how the spectral resolution can influence the physical trend in the mode structure. 

\section{Physical parameter scans} \label{sec:iota_q_scan}

\subsection{Dependence of mode amplitude on $q_\text{edge}$}\label{sec:q_scan}

In this section, VMEC computations are used to carry out parameter scans over the magnetic geometry defined by $q(s)$ and $\iota_\text{ext}$. Initially, a scan of $q_\text{edge}$ was carried out in the tokamak case. \textcolor{black}{To prescribe the initial equilibria for this scan, $q_\text{edge}$ is varied by modifying the toroidal flux of the unperturbed equilibrium, keeping the same total toroidal plasma current and initial current density profile.}

\begin{figure}
    \centering
    \includegraphics[width=0.45\textwidth]{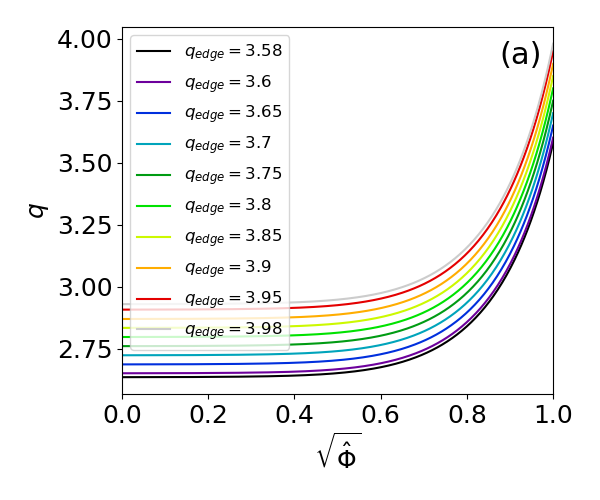}
    \includegraphics[width=0.45\textwidth]{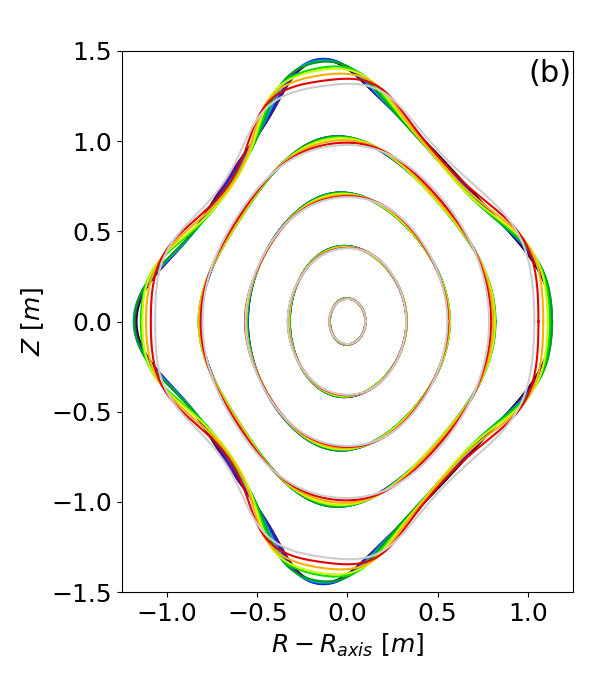}
    \caption{Plots of the $q$ profile (a), and perturbed equilibrium flux surfaces in the $\phi=0$ plane (b) \textcolor{black}{for tokamak cases}. The initial equilibria are computed  with $p=3$ increasing the toroidal flux to modify $q_\text{edge}$. \textcolor{black}{Helicity preserving, free boundary computations are then used to compute the perturbed equilibria, constraining the $q$ profile.}}
    \label{fig:tok_ext_scan_surfaces}
\end{figure}

The $q$ profiles are shown in Figure \ref{fig:tok_ext_scan_surfaces} (a). As expected for a high aspect ratio tokamak case, where $B_{\phi}$ is approximately uniform, the increase of the toroidal flux does not significantly distort the shape of the $q$ profile, only shifting it upwards. The flux surfaces of the perturbed solution in the $\phi=0$ plane are shown in Figure \ref{fig:tok_ext_scan_surfaces} (b). It can be seen that as $q_\text{edge}$ approaches 4, the perturbed mode becomes visibly milder.

\begin{figure}
    \centering
    \includegraphics[width=0.49\textwidth]{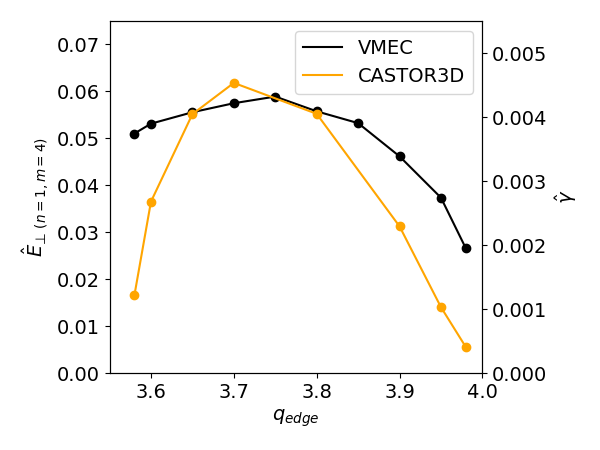}
    
    \caption{Scan of $q_\text{edge}$ for tokamak cases. The $\iota$ profile was preserved through the computations. The overall amplitude of the $(4,\ 1)$ component of the perturbation, $\hat E_{\bot (n=1,\ m=4)}$ (black, left axis), follows the normalised linear growth rate from CASTOR3D (orange, right axis).}
    \label{fig:tok_ext_scan}
\end{figure}

The amplitude of the dominant (4, 1) perturbation is shown in Figure \ref{fig:tok_ext_scan}, alongside the normalised linear growth rate of the (4, 1) external kink mode. According to nonlinear analytic theory, the saturated amplitude of the mode should follow the square of the linear growth rate \cite{eriksson1997nonlinear}. This trend is not necessarily expected across the parameter space considered, because the nonlinear analytic result assumes that the mode is marginally unstable.  Similar to previous studies\cite{kleiner2019current}, it can be seen in Figure \ref{fig:tok_ext_scan} that the general trend of the perturbation amplitude and the linear growth rate are correlated.

It should be noted that computations were also carried out at $q_\text{edge}=3.55$. At this point in the parameter scan, the ideal MHD mode is stabilised. Only a (3, 1) resistive tearing mode was observed in CASTOR3D. It therefore makes sense that a corresponding saturated (4, 1) mode could not be found at $q_\text{edge}=3.55$. As expected, perturbed states were only found in the window of linear instability of the (4, 1) external kink.

\subsection{Dependence of mode amplitude on $\iota_\text{ext}$} \label{sec:iota_scan}

In Figure \ref{fig:tok_ext_scan}, the nonlinear amplitude of the external kink varies most around $q_\text{edge}=3.9$. This value of $q_\text{edge}$ is therefore chosen for scans of $l=2$ stellarators with increasing external rotational transform, in order to observe the influence of $\iota_\text{ext}$ on the saturated mode amplitude. In this scan, $q_\text{edge}$, and $j_0(s)$ are once again held constant, while varying the external rotational transform. 

If the external rotational transform is increased by increasing the $l=2$ deformation of the equilibrium, keeping other equilibrium parameters fixed, $q_\text{edge}$ will decrease. In order to keep $q_\text{edge}=3.9$, the toroidal flux is increased to counteract this effect. The scale invariance arguments invoked in Section \ref{sec:vmec_equilibra} are used to make the time scale, mode amplitude, and nonlinear energy spectra of the results comparable. This can be done by normalising these results by the Alfvén time, minor radius, and square of the magnetic field strength, respectively. 

\begin{figure}
    \centering
    \begin{minipage}{0.45\textwidth}
    \centering
      \includegraphics[width=\textwidth]{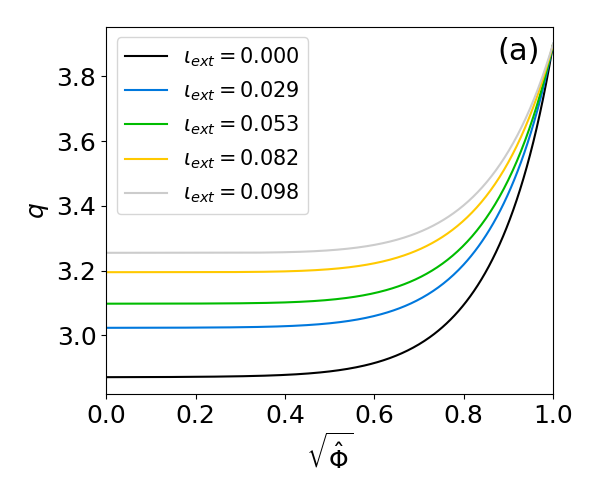}
    \end{minipage}
    \begin{minipage}{0.49\textwidth}
    \centering
      \includegraphics[width=\textwidth]{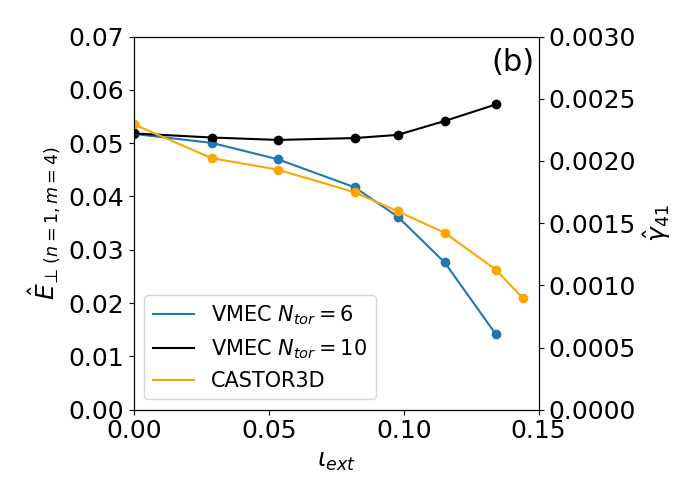}
    \end{minipage}
    \caption{Modification of the $q$ profile (a) with increasing $\iota_\text{ext}$ for $l=2$ $N_\text{p}=2$ stellarators. The behaviour of the nonlinearly saturated perturbation amplitude (b) with increasing $\iota_\text{ext}$ depends on the maximum toroidal mode number used in the computation. The results with $N_\text{tor}=6$ (blue, left y-axis) are likely to be more physical than with $N_\text{tor}=10$ (black,  left y-axis), as the behaviour is correlated with the change in the linear growth rate (orange, right y-axis).}
    \label{fig:iota_ext_l2_Np2_kink_scan}
\end{figure}

\begin{figure}
    \centering
    \begin{minipage}{0.45\textwidth}
    \centering
      \includegraphics[width=\textwidth]{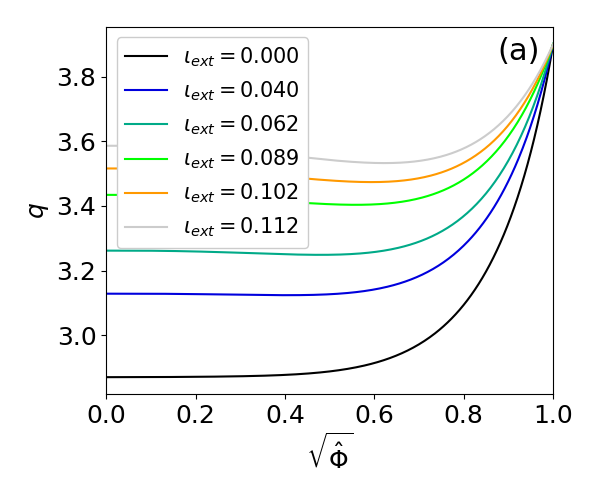}
    \end{minipage}
    \begin{minipage}{0.49\textwidth}
    \centering
      \includegraphics[width=\textwidth]{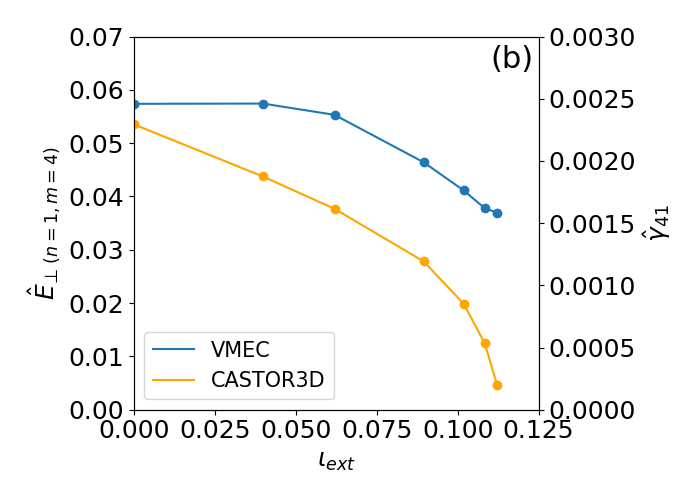}
    \end{minipage}
    \caption{Modification of the $q$ profile (a) with increasing $\iota_\text{ext}$ for $l=2$ $N_\text{p}=5$ stellarators. In (b) it is shown that the nonlinearly saturated perturbation amplitude (blue, left y-axis) decreases with the linear growth rate (orange, right y-axis). Note that computations at higher spectral resolution were not possible for this case, due to the demanding computational requirements.}
    \label{fig:iota_ext_l2_Np5_kink_scan}
\end{figure}

The results of scans of $\iota_\text{ext}$ for $N_p=2$ and 5 are shown in Figure \ref{fig:iota_ext_l2_Np2_kink_scan} and \ref{fig:iota_ext_l2_Np5_kink_scan}, respectively. Once again, the mode amplitudes are compared with the normalised linear growth rates from CASTOR3D. It can be seen that in both cases, a family of saturated states is found, which is incrementally stabilised by the externally imposed helical field. These states are again observed up to the point at which the (4, 1) mode is observed in the linear stability analysis, which again shows that the linear and nonlinear picture presented by the MHD codes used is, in this sense, consistent.

For the $N_p=2$ case, a second branch of saturated states were found when using a higher toroidal resolution in VMEC computations. This second series of equilibria have a larger mode amplitude that does not decrease with the linear growth rate. Comparing computations with $N_\text{tor}=10$ and 12, the solutions obtained were similar, as can be expected from the toroidal mode number resolution scan in Figure \ref{fig:vmec_kink_resolution_scan} (b). 

Because the nonlinear displacement amplitude does not decrease with the linear growth rate, as expected by the nonlinear analytic theory of kink modes \cite{eriksson1997nonlinear}, this second branch is an unexpected result. It is thought that the $N_\text{tor}=10$ states are a second series of bifurcated states which do not correspond to the (4, 1) linear instability. \textcolor{black}{Nevertheless, numerical artifacts could not be identified in the mode structure of these large amplitude solutions, such that they appear to be physical and cannot be discarded conclusively.}

\textcolor{black}{For this reason, it must be acknowledged that even after applying the physical constraints in Section \ref{sec:res_scans}, it was not possible to isolate the physical trend for the dependence on $\iota_\text{ext}$ using VMEC computations alone. In the following sections, aspects of the energy dynamics implied by the VMEC solutions are interrogated to test the physical validity of the perturbed solutions further.}

\section{Energy dynamics and mode coupling of VMEC solutions} \label{sec:nonlinear_energies}

The change in the poloidal magnetic energy spectrum can in principle be used to infer how the magnetic configuration has been modified by the instability, and confirm whether a more energetically favourable state has been reached. It should be noted however that for a free boundary mode, the magnetic energy should be considered over all real space to account for the modification of the energy in the vacuum region. 

Previous studies have understandably only been able to consider the potential energy over the plasma volume covered by the VMEC computational domain \cite{cooper2018stellarator}. While a method to compute the total energy over all space could not be found, the EXTENDER \cite{drevlak2005pies}, NESCOIL \cite{merkel1987solution}, and MAKEGRID codes could be used to compute the plasma and vacuum magnetic fields in a finite domain both inside and outside the plasma volume on a $R$, $Z$, $\phi$ grid. This enables an assessment of the modification of the magnetic field structure in the vacuum region close to the plasma, where the interaction should be strongest.

\begin{figure*}
    \centering
    
    \begin{minipage}{0.325\textwidth}
    \centering
    \includegraphics[width=\textwidth]{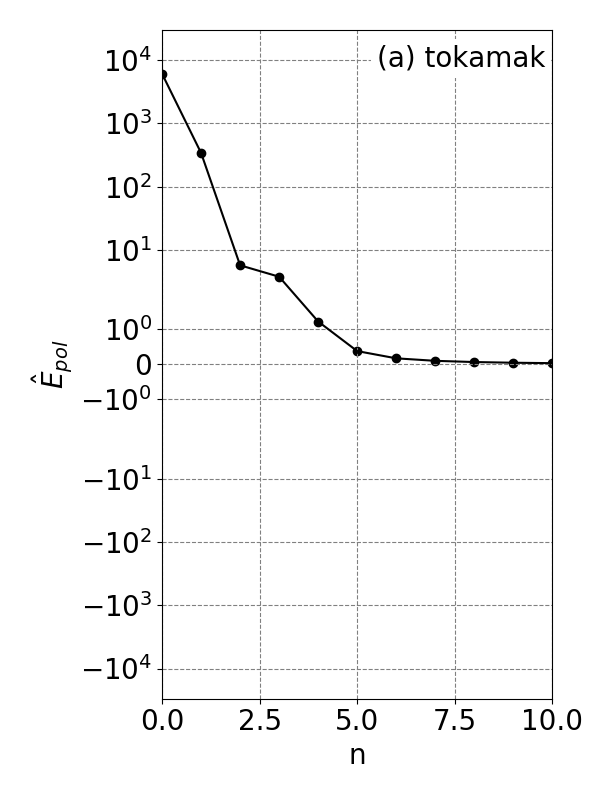}
    \end{minipage}
    \begin{minipage}{0.325\textwidth}
    \centering
    \includegraphics[width=\textwidth]{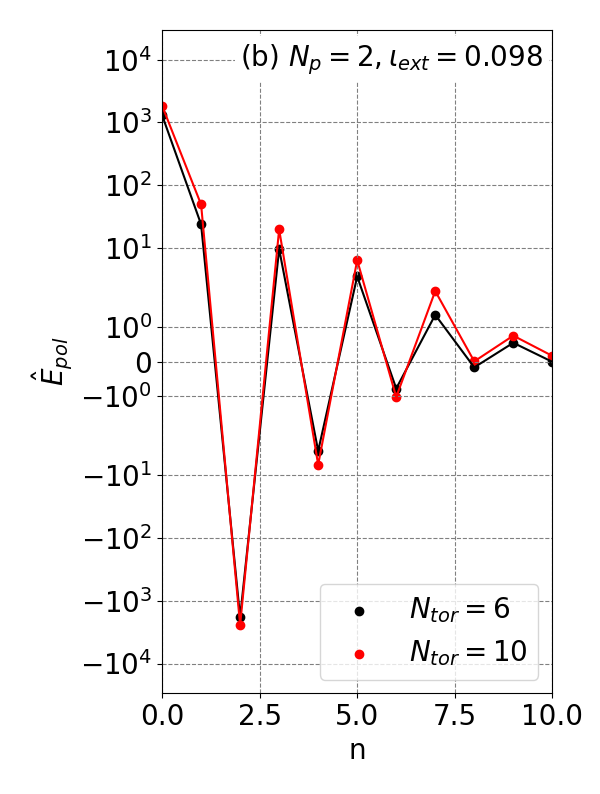}
    \end{minipage}
    \begin{minipage}{0.325\textwidth}
    \centering
    \includegraphics[width=\textwidth]{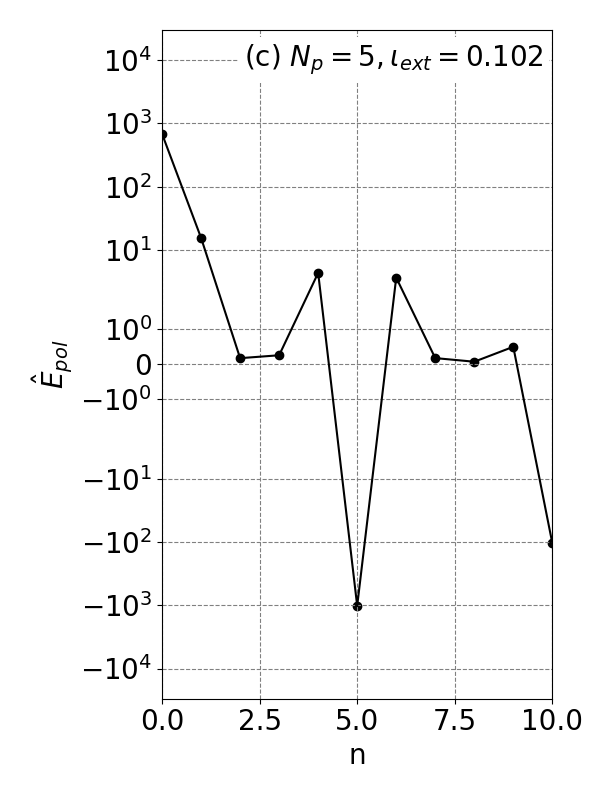}
    \end{minipage}
    \caption{Normalised perturbed poloidal magnetic energy, $\hat E_\text{pol}$, of the tokamak (a), and $l=2$ classical stellarators with $N_\text{p}=2$ (b) and $N_\text{p}=5$ (c) at $q_\text{edge}=3.9$. The energies are integrated over a circular torus, centred at $R=10.0$, with radius, $r_\text{minor}=1.75$, which fully encloses the plasma volume of all cases. $\hat E_\text{pol}$ is normalised by the volume averaged magnetic field strength squared from the unperturbed equilibrium, such that the three cases can be compared quantitatively.}
    \label{fig:q3.9_perturbed_energies}
\end{figure*}

The magnetic field is computed and Fourier decomposed inside a cylindrical torus, centred at $\hat R=1$, with minor radius, $\hat r_\text{minor}=1.75$, where these parameters are normalised by the major and minor radius, respectively. This torus contains all of the simulated cases considered in the parameter scans from Section \ref{sec:iota_q_scan}. The poloidal magnetic energy spectrum is then computed, normalising by the square of the volume averaged toroidal magnetic field such that results of different equilibrium computations can be compared with one another. The results for three cases with different field periodicities are shown in Figure \ref{fig:q3.9_perturbed_energies}. The external rotational transform is $\approx 0.1$ for the stellarator cases in Figure \ref{fig:q3.9_perturbed_energies} (b) and (c). It can be observed that the energy in the perturbation is notably smaller in the stellarator cases. 

In the tokamak case, the magnetic energy cascades from high values in low toroidal harmonics to lower values as $n$ increases. \textcolor{black}{For stellarators however, it can be seen that the perturbation is influenced by the toroidal mode coupling of the simulated device, such that there is distinct behaviour in the different mode families\cite{schwab1993ideal}}. In particular, the perturbation exists predominantly in the $N_\text{f}=1$ mode family. This means that the energy corresponding to the perturbation cascades down from the dominant $n=1$ mode to $n=3$, 5, and so on in the $N_\text{p}=2$ case. Equally in the $N_\text{p}=5$ case, the $n=4$ and $n=6$ modes are notably larger than the members of the $N_\text{f}=2$ mode family. These results are expected due to the toroidal mode coupling of the instability in stellarators. 

In all cases, it can be seen that the $n=0$ poloidal magnetic energy has increased. This result is unintuitive, because the overall energy of the plasma-vacuum system should decrease as a result of the MHD perturbation. The $n=0$ energy provides the drive for the MHD instability, such that the overall energy in this mode is expected to decrease as a result of the nonlinear perturbation, despite the increase in plasma current, observed in Figure \ref{fig:vmec_kink_resolution_scan} (d-f). This unexpected result can be justified by studying the exchange of energy between the plasma and vacuum as considered in Section \ref{sec:n0_pol_mode_structure}.

In contrast, the $n>0$ toroidal harmonics of the $N_\text{f}=0$ mode family show a decrease in poloidal magnetic energy for the $N_\text{p}=2$ and the $N_\text{p}=5$ stellarator results. It could be argued that these toroidal harmonics contribute to the equilibrium drive for the instability, which releases energy from these toroidal harmonics in the perturbed state.

\section{Comparison with nonlinear JOREK simulations} \label{sec:jorek_comp}
A question has been introduced by the observations in Section \ref{sec:nonlinear_energies} - why does the $n=0$ energy increase in all cases? In order to develop an answer to this question, the VMEC results for the tokamak case with $q_\text{edge}=3.9$ are compared with nonlinear simulations using the JOREK code\cite{hoelzl2021jorek}. 

\textcolor{black}{The physics model used in this study to simulate a tokamak is the same as outlined in Ref. \onlinecite{ramasamy2022}, neglecting the parallel momentum equation. However, unlike in this previous study, free boundary conditions are applied to all toroidal harmonics, including the $n=0$ mode, using the JOREK-STARWALL coupling \cite{artola2018free}. In such a way, the $n=0$ energy dynamics can be interrogated, comparing with VMEC. While JOREK has recently been extended to model stellarators\cite{nikulsin2022jorek3d}, the above question of interest can be answered with tokamak simulations alone.} In Section \ref{sec:jorek_sim_setup}, the simulation setup for the tokamak simulation is outlined. The increase in the $n=0$ magnetic energy is then studied in Section \ref{sec:n0_pol_mode_structure}.

\subsection{JOREK simulation parameters} \label{sec:jorek_sim_setup}

The numerical and diffusive parameters used in the simulations are shown in Table \ref{tab:jorek_params}, along with their profiles. The temperature and density linearly decrease from the plasma core to boundary. A smoothing function is used to transition to the vacuum quantities. A Spitzer-like resistivity is used to ensure the vacuum region is highly resistive. The perpendicular diffusive coefficients are kept constant within the plasma region, before artificially being increased outside the plasma using a sigmoid function. This keeps the density and temperature relatively low in the vacuum region. The parallel thermal diffusion coefficient is prescribed as a constant for simplicity. A constant parallel particle diffusivity is also applied to replace the parallel particle transport from advection.

\begin{table}
    \caption{Parameters used in nonlinear free boundary JOREK simulations of a tokamak with $q_\text{edge}=3.9$. The diffusive parameters are the values defined at the plasma core. Unless stated, values are in JOREK normalised units.}
    \begin{ruledtabular}
    \begin{tabular}{@{}c|c|c}
    Parameter & Value & Profile \\ \hline
            $T\ [eV]$                                           &  $0.216 - 648.4$              &  Linear          \\
            $n\ [\times 10^{20}]$                               &  $0.03 - 1.0$                 &  Linear          \\   
            $\kappa_{\parallel}$                                &  $1.0$                        &  Constant     \\
            $\kappa_{\perp}$                                    & $10^{-6}$  &  Sigmoid function    \\ 
            $D_{\parallel}$                                     &  $0.01$                       &  Constant     \\
            $D_{\perp}$                                         & $10^{-6}$  &  Sigmoid function    \\
            $\eta\ [\Omega\cdot m]$                             & $1.9382\times 10^{-7}$        &  Spitzer      \\
            $\eta_{\mathrm{num}}\ [\Omega \cdot {m^3}]$         & $1.9382\times 10^{-12}$       &  Constant     \\
            $\mu\ [kg\cdot m^{-1}\cdot s^{-1}]$                 & $5.1594\times 10^{-8}$        &  Spitzer-like \\
            $\mu_{\mathrm{num}}\ [kg {m} \cdot s^{-1}]$         &  $5.1594\times 10^{-13}$      &  Constant      \\
            $n_{\mathrm{rad}}$                                  &  $143$                        &  --- \\
            $n_{\mathrm{pol}}$                                  &  $86$                         &  --- \\
            $n_{\mathrm{plane}}$                                &  $32$                         &  ---  
    \label{tab:jorek_params}
    \end{tabular}
    \end{ruledtabular}
\end{table}

Typically in nonlinear MHD simulations, an initial phase is required before the onset of the MHD instability, in order to obtain stationary profiles. For the simulated tokamak case, the steady state profiles could not be computed, because the elliptically shaped tokamak equilibrium is vertically unstable. The vertical displacement event would evolve on a much faster time scale than it would take for the equilibrium to reach steady state. For this case, it is necessary to initialise the $n>0$ modes before the profiles have fully equilibrated. For the aspects of the initial ideal saturation that we are interested in, namely the flows of magnetic energy on the fast ideal MHD timescale, the much slower relaxation of the profiles is not considered to have an effect.

\begin{figure*}
    \begin{minipage}{0.33\textwidth}
      \centering
      \includegraphics[width=\textwidth]{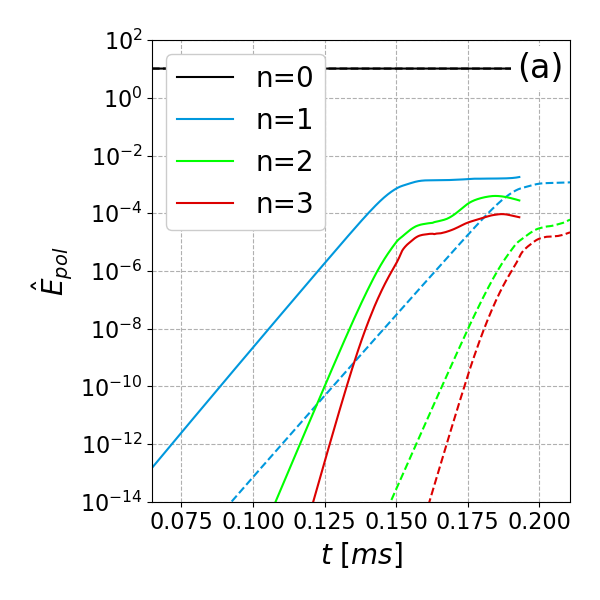}
    \end{minipage}
    \begin{minipage}{0.33\textwidth}
      \centering
      \includegraphics[width=\textwidth]{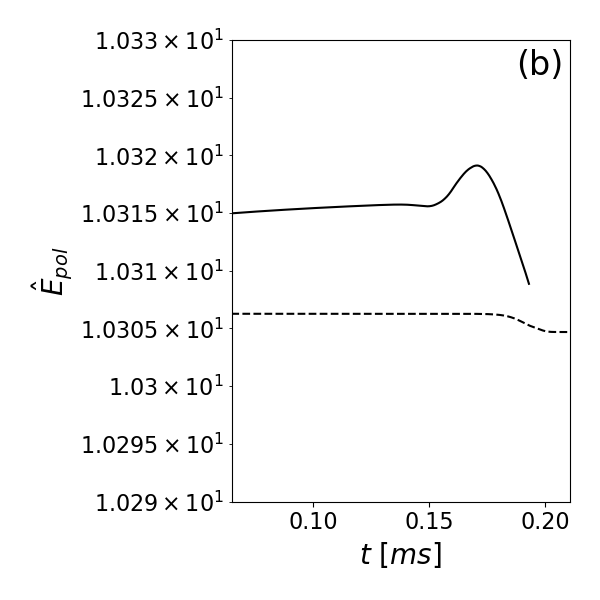}
    \end{minipage}
    \begin{minipage}{0.33\textwidth}
      \centering
      \includegraphics[width=\textwidth]{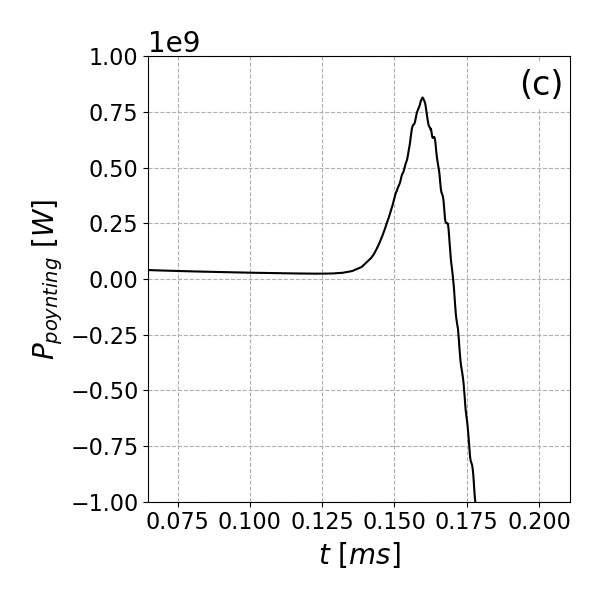}
    \end{minipage}
    \caption{Evolution of poloidal magnetic energy spectrum (a) in $n=0$ free (solid) and $n=0$ fixed (dashed) boundary JOREK simulations of the tokamak case with $q_\text{edge}=3.9$. \textcolor{black}{Considering the $n=0$ energy alone (b), it can be seen that there is an increase in energy in the free boundary case.} The Poynting flux (c) into the free boundary JOREK simulation domain indicates the increase in $n=0$ integrated poloidal magnetic energy is produced by a flux of energy into the simulation domain from the vacuum region.}
    \label{fig:tok_q3.9_freeb_jorek_energies}
\end{figure*}

\subsection{Understanding the increase in $n=0$ poloidal magnetic energy} \label{sec:n0_pol_mode_structure}

To further develop an understanding of the $n=0$ poloidal magnetic energy increase observed in the corresponding VMEC result, the evolution of the poloidal magnetic energies in the JOREK simulation is shown in Figure \ref{fig:tok_q3.9_freeb_jorek_energies} (a). It can be seen in Figure \ref{fig:tok_q3.9_freeb_jorek_energies} (b) that as the $n=1$ external kink saturates, an increase in the $n=0$ energy is observed, similar to in VMEC computations. 

To confirm that this observation is related to the finite integral over the simulation domain, which does not capture all of the energy in the free boundary system, a simulation is also run applying a fixed boundary condition to the $n=0$ poloidal flux. The energies from this simulation are also plotted with dashed lines in Figure \ref{fig:tok_q3.9_freeb_jorek_energies} (a) and (b). With this boundary condition, the $n=0$ energy does not increase. The result is expected because the fixed boundary implies that the simulated domain is a closed system, such that the total magnetic energy cannot increase as a result of the instability. Finally, the Poynting flux on the JOREK boundary is plotted for the free boundary simulation case in Figure \ref{fig:tok_q3.9_freeb_jorek_energies} (c). This diagnostic shows that there is a flux of magnetic energy into the system from outside of the JOREK computational domain. 

The fixed boundary simulation, and the Poynting flux diagnostic confirm that the increase in the $n=0$ energy in JOREK is due to an exchange of energy between the integrated volume, and the vacuum region. While equivalent diagnostics could not be identified to confirm this for the VMEC result, it is likely that the increase in the $n=0$ magnetic energy in Figure \ref{fig:q3.9_perturbed_energies} is a result of the same effect.
 
The exchange of energy is likely due to the changing current profile, and total plasma current. To first order, the confined plasma can be treated as a current carrying loop. If such a current loop has an increasing current, similar to the spike in current observed in the VMEC result, the induced Poynting flux will be towards the current loop. While the rearrangement of current in VMEC and JOREK results is more complex than such a simplified case, the first order effect of the \textcolor{black}{transition from the initial unperturbed equilibrium to saturated state} should be similar, explaining the increase in the $n=0$ energy in the integrated region. This analogy breaks down when considering the energy over all space, which would increase in the case of a loop with increasing current. In the case of the external kink, it is expected that the total energy would decrease because of changes in the spatial current distribution, leading to an effective change of the loop inductance.

\section{Conclusion} \label{sec:conclusion}

\textcolor{black}{The aim of this work was to determine to what extent free boundary VMEC computations could be used to study external kinks in tokamaks and classical stellarators. The authors hoped to provide an initial understanding of how the nonlinear mode structure depends on the magnetic geometry, independent of other nonlinear methods, or experimental data.} 

It has been shown with numerical resolution scans that helicity preserving, free boundary computations do not converge to a single perturbed solution with increasing spectral resolution. As such, the numerical resolution needs to be chosen carefully in order to obtain a physically meaningful result. In this work, additional physical constraints, using the observed plasma current spike and conservation of volume are used to constrain the spectral resolution. \textcolor{black}{Even after applying these constraints to narrow down the range of physically reasonable numerical resolution parameters, it was not possible to identify the physical dependency of the mode amplitude on the external rotational transform using VMEC alone.} For $l=2$, $N_p=2$ stellarators, the trend differs depending on the number of toroidal harmonics included in the VMEC solution.

\textcolor{black}{Despite this limitation, several aspects of the energy dynamics implied by the VMEC results appear physical.} Considering the magnetic energy spectrum in more detail, the perturbed energy is carried predominantly in the $N_f=1$ mode family of stellarator cases, as expected due to toroidal mode coupling. Comparing with JOREK simulations, it is argued that the increase in the $n=0$ magnetic energy observed in VMEC computations is due to an exchange of energy between the plasma and vacuum region.

There are several directions for future work. \textcolor{black}{With respect to the applicability of VMEC to modelling nonlinear MHD, the physical constraints applied in the current work were not sufficient to provide a conclusive answer to the physics questions of interest. It is entirely possible that additional constraints could improve the current situation.} The formation of current sheets at the plasma boundary in Section \ref{sec:current_sheets} could be a potential direction for further constraining the VMEC solution space. If the structure of the expected current sheets at the boundary can be known a priori, perhaps by comparing with linear theory, it may be possible to constrain the VMEC solutions further. The authors personally believe this is a promising constraint to explore. 

It would also be interesting to see if the physical conservation laws and constraints considered herein could be enforced in VMEC during the computation itself. VMEC was afterall not designed for the capture of nonlinearly saturated MHD perturbations. It could be possible that modifying the convergence algorithm could improve the physical validity of the solutions, without having to filter through the solution space manually, as has been done in this work. Such an endeavour has not been attempted thus far, to the author's knowledge. 

Further work could also include additional physics studies, which apply the methods in this paper to consider several open advanced stellarator research questions. Low-n external modes observed on the W7-AS stellarator \cite{merkel1996free}, or external modes predicted in candidate quasi-axisymmetric stellarators \cite{Strumberger2019} could be considered. Such VMEC computations would help to gain an initial understanding of the nonlinear \textcolor{black}{saturation of these MHD modes} to help inform more detailed nonlinear studies with initial value codes, which could include non-ideal, and extended MHD effects. Of course, dedicated nonlinear studies using the recent stellarator extension implemented in JOREK will also be important to interrogate such physics problems.

\section*{Acknowledgements}
The authors would like to thank Florian Hindenlang, Ksenia Aleynikova, Nikita Nikulsin, Guillermo Su\'arez L\'opez, Michael Drevlak and Carolin Nuehrenberg for helpful discussions and assistance in the use of the codes used in this work. 

Some of this work was carried out on the high performance computing architectures COBRA and RAVEN operated by MPCDF in Germany, JFRS-1 operated by IFERC-CSC in Japan, and the EUROfusion High Performance Computer (Marconi-Fusion).

This work has been supported in part by the Max-Planck/Princeton Center for Plasma Physics, and has been carried out within the framework of the EUROfusion Consortium, funded by the European Union via the Euratom Research and Training Programme (Grant Agreement No 101052200 — EUROfusion). Views and opinions expressed are however those of the author(s) only and do not necessarily reflect those of the European Union or the European Commission. Neither the European Union nor the European Commission can be held responsible for them.

\appendix

\begin{figure}
    \centering
    \includegraphics[width=0.45\textwidth]{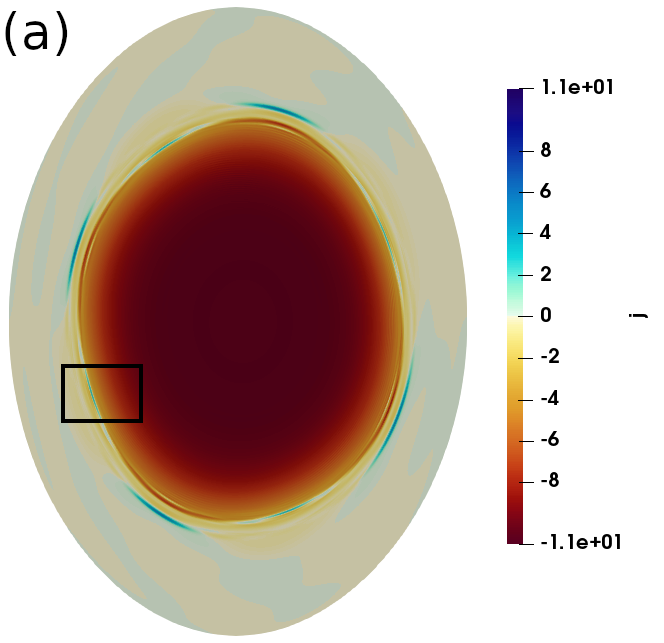}
    \includegraphics[width=0.45\textwidth]{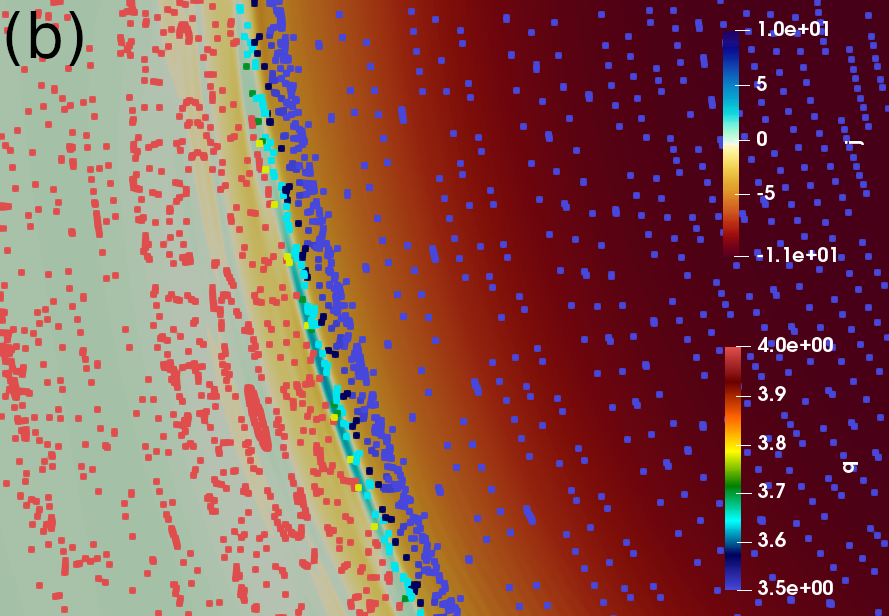}
  
    \caption{The current density $\jmath=Rj_\phi$ in the $\phi=0$ poloidal plane at $t=0.150\ \text{ms}$ (a) of the free boundary tokamak simulation in JOREK. \textcolor{black}{Zooming into the black rectangle (b), negative current sheets are present near the plasma boundary.} The overlaid Poincar\'e data is labelled by the approximate q value of each line. The current sheets are found inside the plasma region where $q < q_\text{edge} = 3.9$.}
    \label{fig:tok_q3.9_freeb_jorek_current}
\end{figure}

\begin{figure}
    \centering
    \includegraphics[width=0.49\textwidth]{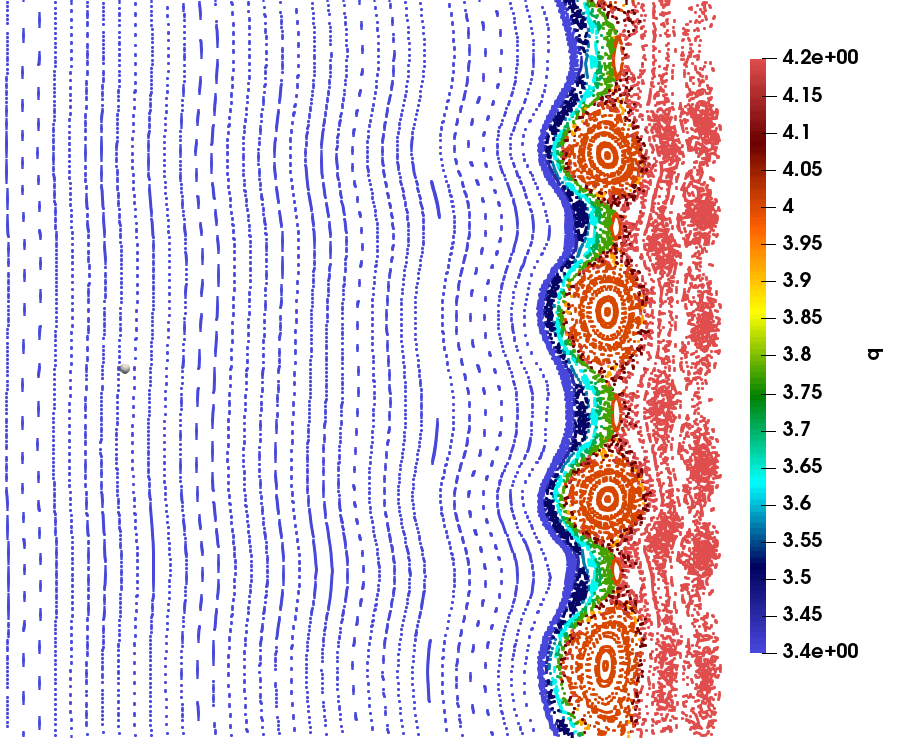}
  
    \caption{Poincar\'e plot at $t=0.161\ \text{ms}$ for the free boundary tokamak case in Figure \ref{fig:tok_q3.9_freeb_jorek_energies} (a). Field lines are labelled by their approximate q value. A secondary (7, 2) island structure (dark blue) has formed, and there is a competing (8, 2) island structure outside the plasma.}
    \label{fig:tok_q3.9_freeb_jorek_poincare}
\end{figure}

\section{Comparison of current sheets in JOREK and VMEC} \label{sec:jorek_currents}

The toroidal current density at $t=0.150\ \text{ms}$ during the initial saturation of the $q_\text{edge}=3.9$ tokamak case studied in Section \ref{sec:n0_pol_mode_structure} is plotted in Figure \ref{fig:tok_q3.9_freeb_jorek_current}. This is the same case that is used to study the current sheets induced in VMEC in Section \ref{sec:current_sheets}. In Figure \ref{fig:tok_q3.9_freeb_jorek_current} (a), broad counterflowing current sheets are observed in the JOREK solution near the x-points of the (4, 1) magnetic islands outside the plasma, as well as much finer current sheets in the hotter plasma edge region, where the kink displacement is radially inwards.

Figure \ref{fig:tok_q3.9_freeb_jorek_current} (b) shows a zoom in of the black rectangle in Figure \ref{fig:tok_q3.9_freeb_jorek_current} (a). \textcolor{black}{Poincar\'e} data is overlaid on the plot, with each field line coloured based on its $q$ value, computed numerically by tracing the field for 200 toroidal turns. In such a way, points with $q < q_\text{edge} = 3.9$ identify the approximate plasma region. It can be seen that the current sheet lies within the plasma volume, intersecting with field lines in the range of $q=3.65-3.8$. The above observations indicate that the (4, 1) current sheets in the VMEC result in Figure \ref{fig:vmec_tok_fixed_radial_scan_current_sheets} (a) are consistent with JOREK in the early nonlinear phase.

Later in the nonlinear phase, competing mode structures are observed in the JOREK solution. For example, (7, 2) and (8, 2) island structures can be observed in Poincar\'e plots computed at $t=0.161\ \text{ms}$, shown in Figure \ref{fig:tok_q3.9_freeb_jorek_poincare}. As argued in previous work\cite{ramasamy2022}, similar island structures triggered by nonlinear mode competition are suppressed by the assumption of nested flux surfaces in VMEC. The (7, 2) Fourier component observed in Figure \ref{fig:vmec_tok_fixed_radial_scan_current_sheets} (b) could represent shielding currents that prevent such islands from forming.

\section*{References}
\bibliographystyle{iopart-num}
\bibliography{references}

\end{document}